\documentclass[12pt,thmsa]{article}
\usepackage{amssymb}

\usepackage{sw20lart}

\input{tcilatex}
\begin{document}

\author{The Author}
\title{The Title }
\date{The Date }
\maketitle

\begin{abstract}
Replace this text with your own abstract.
\end{abstract}

\begin{center}
Letter to the Editor of \bigskip the Journal of Theoretical Biology

\textbf{Long-Bone Allometry of Terrestrial Mammals and the Geometric-Shape
and Elastic-Force Constraints of Bone Evolution}

.
\end{center}

A natural similarity in body dimensions of terrestrial animals noticed by
ancient philosophers remains the main key to the problem of mammalian
skeletal evolution with body mass explored in theoretical and experimental
biology and tested by comparative zoologists. We discuss the long-standing
problem of mammalian bone allometry commonly studied in terms of the
so-called ''geometric'', ''elastic'', and ''static stress'' similarities by
McMahon (1973, 1975a, 1975b). We revise the fundamental assumptions
underlying these similarities and give new physical insights into
geometric-shape and elastic-force constraints imposed on spatial evolution
of mammalian long bones.

A realistic description of an animal skeletal bone might be thought in terms
of a hollow, irregular, curved structural rigid beam, whose linear
dimensions guarantee the avoidance of fracture caused by axial and non-axial
external peak loads. Meanwhile, most mammalian skeletal bones manifest a 
\emph{geometric similarity} that permits one to introduce a \emph{%
cylinder-shape approximation} through the characteristic dimensions $D_{is}$
and $L_{is}$ related to bone diameter and length for a given bone $i$ of a
certain mammalian species $s$. This surmise can be verified experimentally
with the help of bone allometry, which provides data on mammalian allometric
exponents $d_{i}$ and $l_{i}$ through the power-law scaling regressions $%
D_{is}=c_{is}M^{d_{i}}$ and $L_{is}=h_{is}M^{l_{i}}$, where \emph{body mass }%
$M$ ($\equiv M_{s}^{(body)}$) is treated as an external \emph{mammalian}
parameter and $c_{is}$ and $h_{is}$ are constants. The $i$-bone allometric
exponents obey the cylindric-shape evolution \emph{constraint }equation $%
2d_{i}+l_{i}=1$. This immediately follows from (i) the relation between 
\emph{bone mass\ } $M_{is}$ and its volume $D_{is}^{2}L_{is}$ ($M_{is}=$ $%
\rho D_{is}^{2}L_{is}$ , $\rho $ is bone density), (ii) the aforegiven
scaling regressions, and from (iii) McMahon's hypothesis that skeletal mass
is linearly scaled to body mass, \textit{i.e.}, $M_{s}^{(skel)}\propto
M_{is}\propto M$. In what follows, we give an analysis of the bone dimension
scaling in terms of the ''overall-bone'' (hereafter effective bone)\emph{\
mammalian} allometric exponents $d=$ $n^{-1}\Sigma _{i=1}^{n}d_{i}$ and $%
l=n^{-1}\Sigma _{i=1}^{n}l_{i}$, which obey the corresponding
cylindric-shape constraint $2d+l=1$ resulting from the corresponding $i$%
-bone constraint averaged over $n$ mammalian bones. Below we give
statistical analysis for experimental data by Christiansen (1999b) on
allometric exponents for mammalian long bones and show that it corroborates
the cylindric-shape approximation very accurately.

More sophisticated bone allometric studies provide information on evolution
of real non-circular hollow bones with a cross section $A_{is}^{(\exp )}$ ($%
\propto M^{a_{i}}$), approximated by coaxial cylinders of length $L_{is}$,
with external $D_{is}$ and internal $d_{is}$ diameters and of area $%
A_{is}=\pi (D_{is}^{2}-d_{is}^{2})/4$. As follows from Table 2 by Selker and
Carter (1989) given for allometric exponents for Artiodactyl long bones, the
circle-cross-section geometric similarity explicit in the constraint $%
a_{i}=2d_{i}$ works well (with the accuracy amounting to $5\%$).
Furthermore, similar to the internal bone diameter, the bone curvature, also
shows a great deal of variation between different $i$-bones in the same
animal and among different $s$-animals, but does not show a significant
influence from animal mass (Selker and Carter, 1989). We see that geometric
similarity does not depend on local deviations of bone shape from the ideal
cylinder form and can therefore be castled into the cylindric-shape
evolution constraint equation. It is worth noting that the geometric
similarity by McMahon (1975b) corresponds to a particular case of the \emph{%
isometric solution }of the cylindric-shape constraint given by $%
d_{0}=l_{0}=1/3$ and $a_{0}=2/3$.

Generally speaking, \emph{elastic-force }mammalian similarity suggests that
there exists a unique mechanism, attributed to the natural elasticity of the
bones of adult animals, that provides an escape from the critical elastic
deformation (fracture) of bones under peak mechanical stresses. This
similarity is ensured by a safety factor of about 2-4, independent of
mammalian mass (see Biewener, 1990 among others). Following Rashevsky (1948)
and McMahon (1973), the elastic similarity can be introduced in explicit
form on the basis of a mechanical analogy that takes place between a given
bone (or an animal trunk) and a quasi-cylindrical, rigid beam. Elastic-force
mammalian similarity, in a way, completes the geometric-shape similarity
that eventually provides predictions for the dimension-growth effective bone
exponents $d$ and $l$ tested by long bone allometry. For the case of long
bones, it seems plausible to approximate the long ($D_{is}\ll L_{is}$)
cylindrical beams by rigid rods for which the condition of elastic
instability against \emph{axial }loads was first established by Euler. This
was given in terms of the critical \emph{elastic-buckling} force $%
F_{buckl}^{(crit)}{=c\pi ^{2}EI_{is}/L_{is}^{2}}$ where $E$ is the elastic
modulus and $I{_{is}}=\pi D{_{is}}^{4}/64$ is the second cross-area moment
of inertia. A numerical factor ${c}$ depends on the boundary conditions, 
\textit{i.e.} on the way of application of the axial external forces at the
rod ends. One can show (see e.g. Landau and Lifshits, 1989) that $c=1$ and $%
4 $ for the cases of both the non-fixed (on the hinges) ends and fixed ends,
respectively. The first case was cited by Hokkanen (1986) concerning with a
clarification of the elastic similarity employed by McMahon and Rashevsky.
Then, Carter and Spengler (1982, cited by Selker and Carter, 1989) reported
that long bones in living animals are extremely rarely fractured by axial
loads. This is due to the curved structure of bones, which in the case of 
\textit{in vivo} axial loading introduce \emph{bending} moments in their
diaphysis, that create stresses of a greater amplitude than those caused by
pure axial loading (Selker and Carter, 1989). This real case can also be
related to Euler's solution with $c=1$. McMahon (1973) extended these
conditions for the case of the limbs of trees, when a free horizontal \emph{%
bending }of the top from the trunk is caused by proper weight. This results
in $c=1/4$ and permits one to summarize all the buckling-deformation cases
for the critical elastic force as

\begin{equation}
F_{elast}^{(\max )}=F_{buckl}^{(crit)}\backsim E\frac{D_{is}^{4}}{L_{is}^{2}}%
.  \label{1}
\end{equation}

McMahon (1973) suggested additionally that all bone elastic forces for
terrestrial mammals are caused solely by animal weight ($F_{buckl}^{(crit)}{%
\thicksim }$ $gM$, $g$ is the gravitation constant) and therefore McMahon's
elastic similarity for mammalian long bones can be introduced (see also
discussion by Hokkanen, 1986) through (i) the elastic-rod, critical-force
and (ii) the cylindric-shape constraints, namely

\begin{equation}
a\text{)}~4d-2l=1~\text{and~}b\text{)}~2d+l=1\text{.}  \label{2}
\end{equation}
Similarly to the aforegiven $b$ constraint, the elastic-force $a$ constraint
in (2) follows from the allometric bone-dimension scaling applied to (1)
further averaged over all long bones. The system of two equations provides
the well known (McMahon, 1973) elastic model predictions for the bone
diameter $d_{0}^{(buckl)}=3/8$, bone length $l_{0}^{(buckl)}=1/4$ exponents,
along with the length-to-diameter ($l_{0}^{(buckl)}/$ $d_{0}^{(buckl)}$)
dimension-bone exponent $\lambda _{0}^{(buckl)}=2/3$, related to the
allometric scaling $L_{is}\backsim D_{is}^{\lambda _{i}}$. The latter was in
part corroborated by bone allometry by McMahon (1975b) who derived $\lambda
_{F}^{(\exp )}=0.67$, $0.52$ and $0.83$ for \emph{families}, respectively,
Bovidae, Suids, and Cervids in the order Artiodactyla. A similar study by
Alexander (1997) established $d_{B}^{(\exp )}$ $=0.34$ and $l_{B}^{(\exp )}$ 
$=0.26$, with $\lambda _{B}^{(\exp )}$ $=0.76$ for the family Bodivie. When
a wider variety of phylogenetically and anatomically distinct samples were
analyzed within a larger body size range, experimental allometric exponents
exhibited rather the isometric scenario ($\lambda _{0}$ $=1$) than the
elastic model behavior predicted by McMahon (1973, 1975a). This
incompatibility with long-bone allometric observations was reported among
others by Alexander \textit{et al.} (1979a) and Biewener (1983).
Furthermore, recent systematic investigations by Christiansen (1999a, 1999b)
suggest that experimental observations of bone dimensions ``support neither
geometric nor elastic similarity, making both questionable as a means of
explaining long-bone scaling in terrestrial mammals''. Moreover, according
to criticism by Economos (1983) shared by Christiansen (1999b), no
satisfactory explanation for any power-law scaling observed in mammalian
allometry can be expected. From the physical point of view, however, there
is no doubt that long bones possess simultaneously cylindrical-type-shape
and elastic-critical-force similarities. We therefore revise the underlying
hypotheses of the elastic model by McMahon (1973) to describe the
elastic-force ($F_{elast}^{(\max )}\propto M$ and $%
F_{buckl}^{(crit)}=F_{elast}^{(\max )}$) and geometric-shape ($%
M_{s}^{(skel)}\backsim M$) similarities in terms of constraints given in (2).

First, under locomotion peak dynamic conditions, skeletal elastic forces of
animals may exceed those driven by static body weight (Hokkanen, 1986), and
therefore the peak stresses in bones are due to muscle contractions, rather
than to gravity (Carter \textit{et al.}, 1980, Biewener, 1982, Rubin and
Lanyon, 1984, Selker and Carter, 1989, and Biewener, 1991). This implies
that the peak elastic forces in (1) should be substituted by those of muscle
subsystem (muscles, tendons and ligaments), which stores and return elastic
energy (Farley \textit{et al}. 1993), \textit{i.e.}, $F_{elast}^{(\max
)}=F_{musc}^{(\max )}$. Furthermore, after Rubin (1984) many studies provide
strong evidence that the maximum muscle strains recorded in rigorous
activities of animals are independent of body mass, and thus muscle-induced
bone stresses are $F_{elast}^{(\max )}/A_{m}\propto M^{0}$, where $A_{m}$ is
the cross-section area of muscle fibers. This, in turn, exhibits scaling to
body mass through the \emph{muscle-area }allometric exponent $a_{m}$ defined
through the scaling law $A_{m}\propto M^{a_{m}}$. As a consequence, the
elastic-buckling-force constraint should be \emph{modified} and changed in
(2) for $4d-2l=a$, with $a=a_{m}$.

Second, in view of the observation by Prange \textit{et al}. (1979) of
nonlinear evolution of skeletal mass with body mass $M$, Hokkanen (1986)
noted that the corresponding McMahon's hypothesis ($M_{s}^{(skel)}\backsim M$%
) makes the $b$ constraint inaccurate in (2). With taking into account the
bone-mass-to-body-mass scaling law $M_{is}=$ $\rho D_{is}^{2}L_{is}\propto
M^{b_{i}}$, one obtains the following\emph{\ modified} cylindrical-shape
bone constraint for terrestrial mammals: $2d_{i}+l_{i}=b_{i}$. Modified in
such a way McMahon's constraint results in $a$) $4d-2l=a$ and $b$) $2d+l=b$
that provides a new prediction due to \emph{elastic-buckling similarity},
namely

\begin{equation}
d^{(buckl)}=\frac{a^{(\exp )}+2b^{(\exp )}}{8}\text{ and }l^{(buckl)}=\frac{%
2b^{(\exp )}-a^{(\exp )}}{4}\text{.}  \label{3}
\end{equation}
This prediction is tested below on the basis of the experimental data on
muscle-area $a^{(\exp )}$ and bone-mass $b^{(\exp )}$ mammalian effective
exponents.

Third, after Alexander \textit{et al.} (1979b) it has been widely recognized
that the external critical loads acting parallel to the axis of the bone
shaft (diaphysis) are often fewer that those applied in the perpendicular
direction, although the corresponding peak stresses are of the same order of
magnitude (see Rubin and Lanyon, 1984). Within the context of theory of
elasticity, this experimental finding can be treated in terms of the
thermodynamic instability of an $i$-$s$-bone approximated by a convex shell
of characteristic size $L_{is}$ with fixed ends subjected to deformation of
flexure $H_{is}$. The free energy of the shell (a difference between the
elastic energy and the work of deformation produced by some uniform external
peak pressure $p^{(crit)}$) shows its instability when $p^{(crit)}%
\thickapprox E(H_{is}^{(\max )}/L_{is})^{2}$ (Pogorelov, 1960). Adopting for
a critical condition of bone fracture $H_{is}^{(\max )}\backsim D_{is}$ one
has

\begin{equation}
p^{(crit)}\thickapprox \frac{F_{buckl}^{(crit)}}{D_{is}^{2}}\thickapprox 
\frac{F_{bend}^{(crit)}}{L_{is}D_{is}}\backsim E\frac{D_{is}^{2}}{L_{is}^{2}}%
\text{.}  \label{4}
\end{equation}
A scaling of the critical elastic-buckling force to bone dimensions, that
follows from (4), is given in (1), and that for the elastic-bending case is

\begin{equation}
F_{elast}^{(\max )}=F_{bend}^{(crit)}\backsim F_{tors}^{(crit)}\backsim E%
\frac{D_{is}^{3}}{L_{is}}\gg F_{buckl}^{(crit)}\text{ for }L_{is}\gg D_{is}%
\text{. }  \label{5}
\end{equation}
Besides the critical force $F_{bend}^{(crit)}$ due to a perpendicular load
that a rigid beam can withstand without breaking that was already cited by
Hokkanen (1986) and by Selker and Carter (1989, see (11) and (4),
respectively), we have included a corresponding torsional force $%
F_{tors}^{(crit)}$. An estimate for the latter immediately follows from the
critical torsional angle of a slightly bent and twirled rod considered by
Landau and Lifshits (1989). For the critical torsional moment along the rod
axis, one therefore has $F_{tors}^{(crit)}D_{is}=8.98EI_{is}/L_{is}$ that
results in the scaling given in (5).

Although the importance of \emph{bending} and \emph{torsional} critical
loads in the production of peak bone stresses is well established (Rubin and
Lanyon, 1982, Biewener, 1982, Biewener and Taylor, 1986, Selker and Carter
1989), we have demonstrated a natural property of elasticity of bones,
similar to the geometric-shape similarity, in a certain way does not depend
on details of the local bone geometry and bone-end boundary conditions, 
\textit{i.e.} all possible elastic-force scaling (similarities) follow from
(4). The elastic-force\emph{\ bending-torsional }scaling given in (5)
provides the corresponding critical-force constraint $3d-l=a$, that in
combination with the modified cylindrical-shape constraint $2d+l=b$ results
in the relevant \emph{bending-torsional} criterium for effective-bone
evolution, namely 
\begin{equation}
d^{(bend)}=\frac{a^{(\exp )}+b^{(\exp )}}{5}\text{ and }l^{(bend)}=\frac{%
3b^{(\exp )}-2a^{(\exp )}}{5}\text{.}  \label{6}
\end{equation}
As seen from (6), within the context of the hypotheses adopted by McMahon
(1973) one has $d_{0}^{(bend)}=2/5,l_{0}^{(bend)}=1/5$, and $\lambda
_{0}^{(bend)}=1/2$ that was denominated as a static stress similarity
(McMahon (1975b). Remarkable that both the elastic-force criteria, given in
(3) and (6) show a consistence with the isometric solution. For this
specific case, the muscle subsystem develops independently ($b_{0}=1$) and
isometrically ($a_{0}=2/3$) and both the critical elastic forces scale to
mass as $F_{0}^{(\max )}\varpropto M^{2/3}$ regardless of the underlying
bone-structure protecting mechanism against critical buckling, bending or
torsion deformations (see similar discussion by Selker and Carter, 1989).

Let us test the predictions for the bone-diameter $d^{(pred)}$ and $%
l^{(pred)}$ growth exponents given in (3) and (6) with those known from the
one-scale (overall small and large mammals) bone allometry and those
available in the literature data on $a^{(\exp )}$ and $b^{(\exp )}$. For the
case of $a$ constraint, an estimate for the overall mammalian data,
including birds, on muscle-area allometric exponent obtained among others by
Alexander (1977), Alexander \textit{et al.}, (1981), and Pollock and
Shadwick (1994), \textit{i.e.}, $a_{m}^{(\exp )}=0.77-0.83$, was proposed by
Garcia (2001). The most recent data on the $i$-bone-mass allometric
mammalian exponents $b_{i}^{(\exp )}$by Christiansen (2002) are presented in
the last column of \textbf{Table 1}, and the relevant long-bone mammalian
exponent $b^{(\exp )}$ ($=\Sigma _{i=1}^{4}b_{i}^{(\exp )}/4$) can be
introduced with accounting for statistical error due to distinct
regressions, approximately, by $b^{(\exp )}=1.0-1.1$. In\textbf{\ Fig.1 }we
give a comparative analysis for the distinct bone-evolution scenarios
predicted by the elastic-force buckling (area 1) and bending-torsional
criteria (area 2), on the basis of (3) and (6), respectively, with the help
of the aforegiven experimental data $a^{(\exp )}=a_{m}^{(\exp )}$ and $%
b^{(\exp )}$. The modified and improved cylindric-shape and elastic-force
constraints are shown by solid lines for the limiting data $a^{(\exp )}$ and 
$b^{(\exp )}$ and the dashed lines and open points correspond to the
simplified constraints by McMahon. In general, one can see that the
long-bone allometry data on mammalian effective (all-bone-averaged)
exponents by Biewener (1983), Bertram and Biewener (1992, reestimated by
Garcia, 2001), and Christiansen (1999a) are close to both the elastic-force
criteria. An exception although should be made for the pioneering data by
Alexander \textit{et al}. (1979a), where dispersion due to phylogenetic
spectrum of terrestrial mammals was not reduced (see discussion by
Christiansen, 1999b).

Besides the effective-bone mammalian evolution observed from Fig.1, we give
a \emph{statistical analysis} of the similarity constraints in Table 1 with
the help of the one-scale\footnote{%
Extended statistical analysis of the long-bone allometric data scaled for
small and large mammals by Christiansen (1999a, 1999b) was given by
Kokshenev (2003).} $i$-bone allometric data on the dimension exponents $%
d_{i}^{(\exp )}$, $l_{i}^{(\exp )}$ and the $i$-bone-mass exponents $%
b_{i}^{(\exp )}$ derived by Christiansen (1999a, 1999b and 2002,
respectively) within four decades of mammalian mass ($1$-$10^{5}kg$) through
the two different one-scale regressions. As seen from Table 1, the
cylindrical-shape approximation for the geometric-shape similarity through
the constraint $2d_{i}^{(\exp )}+$ $l_{i}^{(\exp )}=b_{i}^{(\exp )}$ is
experimentally justified with a good level of accuracy by the data on $%
b_{i}^{(\exp )}$ derived by both the methods for all $n=1,2...6$ long bones,
except for the case of the \emph{ulna}. A high precision is established for
the bone-mass allometric exponent $b^{(\exp )}=1.0$-$1.1$ predicted on the
basis of the cylindrical-shape constraint $b^{(pred)}=2$ $d^{(\exp )}+$ $%
l^{(\exp )}$ within 4-bone averaging, with $b^{(pred)}=1.03$-$1.06$.

The elastic-force similarities are tested through the predicted
effective-bone mammalian exponents $a^{(pred)}$, related to the muscle-area
exponents, found as certain predictions given by the elastic-force buckling
and bending-torsional constraints , respectively: $a^{(buckl)}=4d^{(\exp
)}-2l^{(\exp )}$ and $a^{(bend)}=3d^{(\exp )}-l^{(\exp )}$. As follows from
the analysis elaborated in Table 1, the overall-bone predictions for $%
a^{(buckl)}$ and $a^{(bend)}$ are, respectively, wedged between $0.89$-$0.97$
and $0.81$-$0.87$. With taking into account that $a_{m}^{(\exp )}$ is
limited by $0.83$, we infer that the predicted data based on (i) that the
elastic buckling force given in (1) and those for (ii) long bones \emph{%
fibula} and \emph{ulna }(treated by the \emph{RMA} regression method)
contradict to muscle-area allometry and therefore should be excluded from
the critical-force evolution scenario. The first statement is discussed in
(4) and the second is in line with the notes by Christiansen (1999a, 1999b)
that `` too thin \emph{ulna}'' and greatly reduced \emph{fibula} '' do not
have ''any importance in support of body mass''. In the special case of
carnivore's \emph{ulna}, however, provides a substantial forelimb support
(Christiansen, 1999a).

As follows from analysis given in Table 1, the overall-bone (\emph{LSR}
method) prediction for the muscle-area exponent is $a^{(bend)}=0.81$-$0.83$,
which coincides with the \emph{critical-muscle-force }exponent $%
a_{cm}^{(\exp )}=0.81$-$0.83$ provided by muscle-fiber allometry data by
Pollock and Shadwick (1994) reanalyzed recently by Kokshenev (2003). This
finding implies that the statistical analysis of the elastic-force criteria
based on the equation $F_{elast}^{(\max )}=F_{musc}^{(\max )}$ is related to
the peak-area muscles and should therefore exclude muscle fibers with
relatively small cross-sections. This can be exemplified by \emph{common
digital extensors} which have the lowest ''maximum'' area and exhibit
isotropic evolution with $a_{m}^{(\exp )}\thickapprox a_{0}=2/3$, as shown
in Fig.1 by Kokshenev (2003). Finally, let us estimate the aforementioned
length-to-diameter effective bone exponent $\lambda
^{(bend)}=(3b^{(pred)}-2a^{(pred)})/(b^{(pred)}+a^{(pred)})$ predicted with
4-bone exponents $b^{(pred)}=1.03$-$1.06$ and $a^{(pred)}=0.81$-$0.83$
obtained in Table 1. This provides $\lambda ^{(bend)}=0.77$-$0.83$ that
corroborates the allometry exponent $\lambda ^{(\exp )}=0.781$-$0.810$
derived by Kokshenev (2003) from the relevant data by Christiansen (1999b).

To summarize, we have revised McMahon's elastic similarities that have long
been a controversial subject of intensive study in the last three decades,
especially in the long-bone allometry for terrestrial mammals. The
elastic-force and cylindrical-shape bone evolution constraint equations were
not supported experimentally, and we have therefore reconsidered the basic
hypotheses of the approach by McMahon (1973). As a matter of fact,
exploration of the long bone allometric data was commonly given within the
framework of model simplifications such as (i) the skeletal subsystem of
animals operates separately from the muscle subsystem and (ii) the bone
stresses are induced solely by gravity (Rashevsky, 1948, McMahon, 1973).
These contradict to the observed evolution of bone mass and ignores a role
of muscle contractions in formation of peak skeletal stresses. McMahon's
description of the cylindrical-shape and elastic-force similarities implicit
in the bone dimensional constraints is revised, modified and extended in
view of the up-to-date knowledge on the bone-mass and muscle-area scalings
and on domination of the bending-induced deformations.

Unlike the case of McMahon's (1973) elastic similarity model, the
elastic-force (buckling-deformation) and cylindrical-shape similarities
modified by muscle-area and bone-mass evolution (shown by shaded area 1 in
Fig.1) become closer to almost all cited experimental data. At the same
time, the most accurate data by Christiansen (1999a, 1999b) conflicts with
the modified elastic-buckling bone evolution scenario. Further improvement
of the elastic-force similarity by critical bending and torsional
deformations made it eventually observable (see shaded area 2 in Fig.1).
Statistical analysis of the allometric data by Christiansen (1999b, 2002)
corroborates, at least in the case of the standard \emph{LSR} method, the
proposed description of the geometric-shape and elastic-force mammalian
similarities given in terms of the corresponding constraints within the
cylindrical-shape and elastic-rod model approximations.

The authors are indebted to R. McNeill Alexander for sending the detailed
experimental data. Thanks are due to J. Tyson for valuable criticism, which
improved substantially a presentation of the manuscript. Financial support
by the CNPq (V.B.K.) is also acknowledged.\medskip

\bigskip

KEY\ WORDS:

allometry, long bones, mammals\bigskip \bigskip , critical elastic
deformations.

\begin{center}
Kokshenev V.B.$^{*)}$\bigskip , Silva J.K.L., and Garcia G.J.M.
\end{center}

\textit{Departamento de F\'{i}sica, }

\textit{Universidade Federal de Minas Gerais, }

\textit{Instituto de Ci\^{e}ncias Exatas,}

\textit{Caixa Postal 702, CEP 30123-970, }

\textit{Belo Horizonte, MG, Brazil \medskip }

$^{*)}$valery@fisica.ufmg.br, corresponding author.

\begin{center}
(Submitted 30 November 2002, in revised form 14 March 2003)
\end{center}

\newpage

\begin{center}
{\large REFERENCES}
\end{center}

Alexander, R. M. (1977). Allometry of the limbs of antelopes (Bovidae). 
\textsl{J. Zool.} \textbf{183}, 125-146.

Alexander, R. M., Langman, V.A \& Jayes, A. S. (1977). Fast locomotion of
some African ungulates. \textsl{J. Zool. }\textbf{183}, 291-300.

Alexander, R. M., Jayes, A. S., Maloiy, G. M. O. \& Wathuta, E. M. (1979a).
Allometry of the limb bones of mammals from shrews (Sorex) to elephant
(Loxodonta). \textsl{J. Zool.} \textbf{189}, 305-314.

Alexander, R. M., Maloiy, G.M.O., Hunter B., Jayes, A. S. \& Nturibi J.
(1979b). Mechanical stresses in fast locomotion of buffalo (Syncerus caffer)
and elephant (Loxodonta africana). \textsl{J. Zool.} \textbf{189}, 135-144.

Alexander, R. M., Jayes, A. S., Maloiy, G.M.O. \& Wathuta, E. M. (1981).
Allometry of the leg muscles of mammals, \textsl{J. Zool.} \textbf{194},
539-552.

Bertram, J. E. A. \& Biewener, A. A. (1992). Allometry and curvature in the
long bones of quadrupedal mammals. \textsl{J. Zool.} \textbf{226}, 455-467.

Biewener, A. A. (1982). Locomotory stresses in the limb bones of two small
mammals: the ground squirrel and chipmunk. \textsl{J. Exp. Biol.} \textbf{103%
}, 131-154.

Biewener, A. A. (1983). Allometry of quadrupedal locomotion - the scaling of
duty factor, bone curvature and limb orientation to body size. \textsl{J.
Exp. Biol.} \textbf{105}, 147-171.

Biewener, A. A. \& Taylor , C.R. (1986). Bone strain: a determinant of gait
and speed? \textsl{J. Exp. Biol.} \textbf{123}, 383-400.

Biewener, A. A. (1990). Biomechanics of mammalian terrestrial locomotion. 
\textsl{Science} \textbf{250}, 1097-1103.

Biewener, A. A. (1991). Musculoskeletal design in relation to body size. 
\textsl{J. Biomech.} \textbf{24}, 19-29.

Carter, D.R., Smith, D.J., Spengler, D.M., Daly, C.H.,\& Frankel, V.H.
(1980). Measurement and analysis on \textit{in vivo} bone strains on the
canine radius and ulna. \textsl{J. Biomech.} \textbf{13}, 27-38.

Christiansen, P. (1999a). Scaling of the limb long bones to body mass in
terrestrial mammals. \textsl{J. Morphol.} \textbf{239}, 167-190.

Christiansen, P. (1999b). Scaling of mammalian long bones: small and large
mammals compared. \textsl{J. Zool.} \textbf{247}, 333-348.

Christiansen, P. (2002). Mass allometry of the appendicular skeleton in
terrestrial mammals. \textsl{J. Morphol.} \textbf{251}, 195-209.

Economos, A. C. (1983). Elastic and/or geometric similarity in mammalian
design? \textsl{J. Theor. Biol.} \textbf{103}, 167-172.

Farley C.T., Glasheen J., and McMahon T.A. (1993). Running springs:speed and
animal size. \textsl{J. Exp. Biol.} \textbf{185}, 71-86.

Garcia, G.J.M. (2001). Leis de escala em biologia, \textsl{M.S. thesis},
Universidade Federal of Minas Gerais, unpublished.

Hokkanen, J. E. I. (1986). Notes concerning elastic similarity. \textsl{J.
Theor. Biol.} \textbf{120}, 499-501.

Kokshenev, V.B. (2003). Observation of mammalian similarity through
allometric scaling laws. \textsl{Physica A}, in press.

Landau, L.D. \& Lifshits, E.M. (1989). Theory of elasticity, Pergamon Press,
London.

McMahon, T. A. (1973). Size and shape in biology. \textsl{Science} \textbf{%
179}, 1201-1204.

McMahon, T. A. (1975a). Allometry and biomechanics: limb bones in adult
ungulates. \textsl{Am. Nat.} \textbf{109}, 547-563.

McMahon, T. A. (1975b). Using body size to understand the structural design
of animals: quadrupedal locomotion. \textsl{J. Appl. Physiol. }\textbf{39},
619-627.

Pogorelov, A. V. (1960). Elastic deformations of convex shells in the
transcritical region. \textsl{Dokl. Acad. Nauk SSSR} 133, 785-787.

Pollock, C. M. \& Shadwick, R. E. (1994). Allometry of muscle, tendon, and
elastic energy-storage capacity in mammals. \textsl{Am. J. Physiol.} \textbf{%
266}, R1022-R1031.

Prange, H. D., Anderson, J. F. \& Rahn, H. (1979). Scaling of skeletal mass
to body mass in birds and mammals. \textsl{Am. Nat.} \textbf{113}, 103-122.

Rashevsky, N. (1948). Mathematical Biophysics, revised edition. Chicago: The
University of Chicago Press. (New edition, 1960. New York: Dover
Publications).

Rubin, C. T. \& Lanyon, L. E. (1982). Limb mechanics as a function of speed
and gait: a study of functional strains in the radius and tibia of horse and
dog. \textsl{J. Theor. Biol.} \textbf{101}, 187-211.

Rubin, C. T. \& Lanyon, L. E. (1984). Dynamic strain similarity in
vertebrates; an alternative to allometric limb bone scaling. \textsl{J.
Theor. Biol.} \textbf{107}, 321-327.

Selker, F. \& Carter, D. R. (1989). Scaling of the long bone fracture
strength with animal mass. \textsl{J. Biomech.} \textbf{22}, 1175-1183.

\begin{center}
\newpage 
\begin{tabular}{|l|l|l|l|l|l|l|l|}
\hline\hline
& methods & LSR & LSR & buckl & bend & cylind & cylind \\ \hline\hline
$i$ & Bones & $d_{i}^{(\exp )}$ & $l_{i}^{(\exp )}$ & $\ a_{i}$ & $\ a_{i}$
& $\ b_{i}$ & $b_{i}^{*}$ \\ \hline\hline
1 & humerus & $.3816$ & $.2996$ & $0.93$ & $0.85$ & $1.06$ & $1.07$ \\ \hline
2 & femur & $.3548$ & $.3014$ & $0.82$ & $0.76$ & $1.01$ & $1.06$ \\ \hline
3 & tibia & $.3600$ & $.2571$ & $0.93$ & $0.82$ & $0.98$ & $0.98$ \\ \hline
4 & radius & $.3868$ & $.2995$ & $0.95$ & $0.86$ & $1.07$ & $1.08$ \\ \hline
5 & fibula & $.3379$ & $.2250$ & $0.90$ & $0.79$ & $0.90$ & $\ -$ \\ \hline
6 & ulna & $.3551$ & $.3016$ & $0.82$ & $0.76$ & $1.01$ & $1.08$ \\ 
\hline\hline
$n$ & Averaged & $d^{(\exp )}$ & $l^{(\exp )}$ & $\ a$ & $a$ & $\ \ b$ & $%
b^{*}$ \\ \hline\hline
4 & - bone & $.3708$ & $.2894$ & $0.90$ & \textbf{0.82} & \textbf{1.03} & $%
1.05$ \\ \hline
5 & - bone & $.3642$ & $.2765$ & $0.90$ & \textbf{0.82} & \textbf{1.00} & $\
\ -$ \\ \hline
6 & - bone & $.3627$ & $.2807$ & $0.89$ & \textbf{0.81} & \textbf{1.01} & $\
\ -$ \\ \hline
\end{tabular}
\begin{tabular}{|l|l|l|l|l|l|}
\hline\hline
RMA & RMA & buckl & bend & cylind & cylind \\ \hline\hline
$d_{i}^{(\exp )}$ & $\ l_{i}^{(\exp )}$ & $\ \ a_{i}$ & $\ a_{i}$ & $\ b_{i}$
& $b_{i}^{(\exp )}$ \\ \hline\hline
$.3860$ & $.3109$ & $0.92$ & $0.85$ & $1.08$ & $1.083$ \\ \hline
$.3599$ & $.3089$ & $0.82$ & $0.77$ & $1.03$ & $1.071$ \\ \hline
$.3654$ & $.2767$ & $0.91$ & $0.82$ & $1.01$ & $0.998$ \\ \hline
$.4014$ & $.3210$ & $0.96$ & $0.88$ & $1.12$ & $1.101$ \\ \hline
$.3681$ & $.2430$ & $0.99$ & $0.86$ & $0.98$ & $-$ \\ \hline
$.4579$ & $.3177$ & $1.20$ & $1.06$ & $1.23$ & $1.101$ \\ \hline\hline
$\ d^{(\exp )}$ & $\ l^{(\exp )}$ & $\ a$ & $\ a$ & $\ b$ & $b^{(\exp )}$ \\ 
\hline\hline
$.3782$ & $.3044$ & $0.90$ & \textbf{0.83} & \textbf{1.06} & $1.063$ \\ 
\hline
$.3762$ & $.2921$ & $0.92$ & $0.84$ & $1.04$ & $-$ \\ \hline
$.3898$ & $.2964$ & $0.97$ & $0.87$ & $1.08$ & $-$ \\ \hline
\end{tabular}
\end{center}

.

.

Table 1. Analysis of the bone evolution mechanism through observation of the
mammalian long-bone similarity on the basis of the elastic-force and
cylindrical-shape evolution constraint equations. Experimental data on the
mammalian bone diameter $d_{i}^{(\exp )}$, length $l_{i}^{(\exp )}$
allometric exponents for 6 long bones obtained by the least square
regression (\emph{LSR}) and the reduced major axis (\emph{RMA}) methods are
taken from Table 2 by Christiansen (1999a). The \emph{RMA} data for the
bone-mass allometry exponent $b_{i}^{(\exp )}$($=b_{i}^{(RMA)}$) is taken
from Table 2 by Christiansen (2002). The \emph{LSR} data is estimated here,
approximately, through the relation $b_{i}^{*}=r_{i}^{(RMA)}b_{i}^{(RMA)}$,
where $r_{i}^{(RMA)}$ is the corresponding correlation coefficient by
Christiansen (2002). The muscle-area $a_{i}$ and the bone-mass $b_{i}$
exponents are predicted with the help of the elastic-force \emph{buckling-}%
deformation ($a_{i}=4d_{i}-2l_{i}$) and \emph{bending-}deformation ($%
a_{i}=3d_{i}-l_{i}$) constraint equations, and\emph{\ }cylindrical-shape
constraint equation ($b_{i}=2d_{i}+$ $l_{i}$). The effective $n$-bone
mammalian allometric exponents are given through the mean values of \ $n$
corresponding exponents, \textit{e.g.}, $a=n^{-1}\Sigma _{i=1}^{n}a_{i}$.

\newpage

\begin{center}
{\Large FIGURE CAPTURES}
\end{center}

\bigskip

Fig.1\textbf{.} Mammalian bone-dimension evolution diagram: diameter against
length. \emph{Solid} \emph{lines} are given by the elastic-force $a$%
-constraint \emph{buckling} ($4d-2l=a$) and \emph{bending} ($3d-l=a$)
equations, and for the cylindrical-shape constraint ($2d+l=b$) equation. 
\emph{Dashed lines} - the same for the case of $a=b=1$. \emph{Points}: (%
\emph{open squares}) 0, 1 and 2 correspond, respectively, to the isometric ($%
1/3$, $1/3$), the elastic similarity ($3/8$, $1/4$), and to the stress
similarity ($2/5$, $1/5$) models by McMahon (1975b); \emph{closed circles}
symbolized by A'79, B'83, B'92 and C'99 are due to one-scale,
overall-bone-averaged, least-square-regression data on the mammalian
long-bone allometry by Alexander \textit{et al}. (1979a), Biewener (1983),
Bertram \& Biewener (1992, reestimated by Garcia, 2001), and Christiansen
(1999a), respectively. Statistical error is shown by bars. \emph{Dashes areas%
} $1$ and $2$ correspond to the proposed criteria given by the predictions $%
d^{(buckl)}$, $l^{(buckl)}$ and $d^{(bend)}$, $l^{(bend)}$ in, respectively,
(3) and (6) for the case of the overall muscle areas with $a_{m}^{(\exp
)}=0.77-0.83$ (by Garcia, 2001) and bone masses with $b^{(\exp )}=1.0-1.1$
(from Table 1).

\end{document}